\documentclass[12pt]{iopart}

\usepackage{graphicx}
\usepackage{amssymb} 
\usepackage{tabularx}
\begin{document}
  \title{Tailoring the core electron density in modulation-doped Core-Multi-Shell nanowires}
   \author{Fabrizio~Buscemi$^1$, Miquel~Royo$^{2,3}$, Guido~Goldoni$^{1,4}$, Andrea~Bertoni$^{4}$}
\address{$^1$Department of Physics, Informatics, and Mathematics, University of  Modena and Reggio Emilia, Via Campi 213/A, 41125 Modena, Italy}
\address{$^2$Departament de Qu\'{i}mica F\'{i}sica i Anal\'{i}tica, Universitat Jaume I, Castell\'{o}, Spain}
\address{$^3$Institut de Ci\`{e}ncia de Materials de Barcelona ICMAB-CSIC, Campus de Bellaterra, Barcelona, Spain}
\address{$^4$Centro S3, Istituto Nanoscienze - CNR, Modena, Italy}
\ead{andrea.bertoni@nano.cnr.it}
\begin{abstract} 
 We show how a proper radial modulation of the composition of core-multi-shell nanowires critically enhances the control of the free-carrier density in the high-mobility core with respect to core-single-shell structures, thus overcoming the technological difficulty of fine tuning the remote doping density.
 We calculate the electron population of the different nanowire layers as a function of the doping density and of several geometrical parameters by means of a self-consistent Schr\"odinger-Poisson approach: Free carriers tend to localize in the outer shell and screen the core from the electric field of the dopants.
\end{abstract}
\submitto{\NT}
%\keywords{core-shell nanowires, modulation doping, tubular 2DEG, electrostatic screening}
%\submitto{\NT}

\maketitle

% \pacs{73.63.Nm, 03.67.Hx, 03.65.Ud, 85.35.Be}

\section{Introduction} \label{SP_Introduction} 
  
Semiconductors nanowires (NWs) have the potential to become a key component of next-generation ultrafast electronic nanodevices such as high mobility transistors,~\cite{Cui2003,Viti_NanoRL2012} light-emitting diodes,~\cite{Tomioka2010}, wavelength-controlled lasers~\cite{Qian2008} and photovoltaic cells.~\cite{Czaban2009}  In particular, controlled epitaxial lateral overgrowth enabled the development of radial modulation doping techniques, a crucial step towards high-mobility NW devices~\cite{Funk2013}.

High conductivity GaAs NWs with a large aspect ratio and a diameter of few nanometers have been demonstrated.~\cite{Sladek2010} 
With remote doping, the conductive core of the NW is spatially separated from the dopants, reducing carrier scattering by impurities. 
However, the controlled incorporation  of dopants in NWs is still challenging~\cite{Jadc2014}, with poor control of the free-carrier density, the latter being a critical parameter to calibrate the electronic characteristics of devices. Indeed, a core electron population which is weakly sensitive to dopant concentration is a key feature for the application of NWs in high performance electronics.

In this article, we show that a proper radial modulation of the semiconductor NW with an additional conductive layer, a so-called quantum well tube (QWT), will critically decrease the impact of doping uncertainty on the free-electron density in the high conductive NW core. To analyse the impact of compositional parameters on free-carrier concentration and localization we use quantum self-consistent calculations as well as a semi-classical approximation for comparison. 

\begin{figure}[b]
  \begin{center} 
   \includegraphics[width=\linewidth]{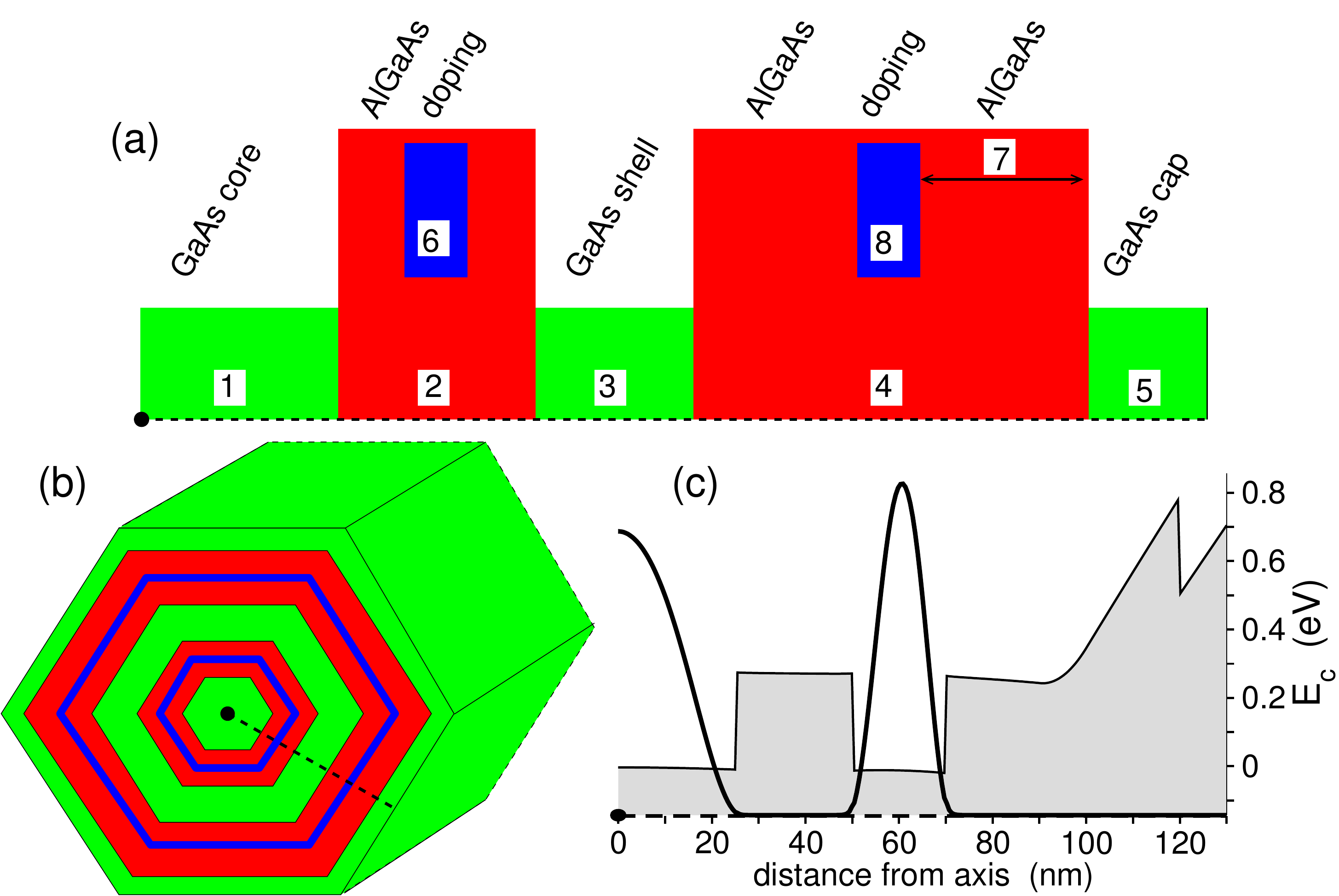}
    \caption{ Schematics of the radial composition (a) and cross-section (b) of a GaAs/AlGaAs MSNW. In (a) composition is represented from the NW center (left) to the outer shell (right). The height of each layer represents $E_C$. Numeric labels refer to layer widths in Tab.~\ref{table1}.
    (c) Example of the self-consistent potential (filled area) and electron density (solid line) along the dashed line of panel (b), for sample $M_1$ (see Tab.~\ref{table1}).
    \label{figure1} }
  \end{center}
\end{figure}

Radial heterostructures have been grown both as core-single-shell NW (SSNW) structures and as core-multi-shell NW (MSNW) structures. In the latter case, one or multiple conductive QWTs can be grown around the NW core. The scheme of an hexagonal MSNW with two insulating AlGaAs shells is shown in Fig.~\ref{figure1}(a) and (b). Different doping levels result in the formation and localization of free carriers gas at the facets, corners or in the center of the hexagonal NW core.~\cite{Bertoni2011,Jadc2014,Morkotter2015,RosdahlNL2015} 
Additionally, in MSNWs carriers can be confined in the QWT wrapped around the core.~\cite{Funk2013} The charge accumulating in the outer shells will eventually screen the electrostatic field generated by remote doping, reducing its effect on the population of the core.

\section{Method} \label{SP_Method} 

Our prototypical heterostructure is a $n$-doped GaAs/AlGaAs MSNWs with hexagonal cross section. However, our results are not limited to this kind of samples since they are not linked to a specific material or symmetry of the NW.
To obtain the free-electron density distributions, we use a self-consistent Schr\"odinger-Poisson approach.~\cite{datta2005, Bertoni2011, Wong2011, Royo2013, Royo2015} 
Within an effective mass, single band approximation, assuming translational invariance along the NW growth $z$-axis, quantum states are given by the 2D Schr\"odinger equation 
\begin{equation}\label{eq_sch}
\left[ -\frac{\hbar^2}{2} \nabla_{\textbf{r}} \frac{1}{m^{\ast}(\textbf{r})} \nabla_{\textbf{r}}  +E_C(\textbf{r})-eV(\textbf{r}) \right]\psi_n(\textbf{r})=E_n\psi_n(\textbf{r}),
\end{equation}
where $\textbf{r}=(x,y)$ is the 2D coordinate, $m^{\ast}$ is the material-dependent effective mass of the electron, $E_C$ and $V$ are the local conduction band edge and the electrostatic potential generated by free electrons and dopants, respectively. Exchange and correlation effects are not included since they are found to be negligible in the devices under consideration.~\cite{Wong2011,Bertoni2011} 
The box integration method on a triangular mesh with hexagonal elements is used to solve Eq.~\ref{eq_sch}~\cite{Bertoni2011, Royo2013} on an hexagonal domain.

Once the wave functions $\psi_n(\textbf{r})$ and subband energies $E_n$ are obtained, the electron density $n(\textbf{r})$ at temperature $T$ is computed from subband population with the Fermi distribution,
\begin{equation}
n(\textbf{r})= 2\sum_n \left|\psi_n(\textbf{r})\right|^2 
\sqrt{\frac{m^{\ast} k_B\textrm{T}}{2\pi \hbar^2}} 
\mathcal{F}_{-\frac{1}{2}}\left(\frac{E_{\textrm{F}}-E_n}{k_B\textrm{T}} \right)
\end{equation}
where 
$E_{\textrm{F}}$ denotes the Fermi level and $\mathcal{F}_{j}(x)=\frac{1}{\Gamma(j+1)}\int_0^{\infty} \frac{t^jdt} {e^{t-x}+1}$ is the complete Fermi-Dirac integral of order $j$ resulting from the integration of the parabolic dispersion along the free axis.

The electrostatic potential $V(\textbf{r})$ due to the electrons and ionized dopants is obtained from the Poisson equation
\begin{equation} \label{eq_poisson}
\nabla_{\textbf{r}} \left[ \epsilon_r(\textbf{r}) \nabla_{\textbf{r}} V(\textbf{r})\right]=-\frac{e}{\epsilon_0} [\rho_D(\textbf{r})-n(\textbf{r})],
\end{equation}
where $\epsilon_r(\textbf{r})$ is the position-dependent dielectric constant, $\epsilon_0$  is the vacuum permittivity, and $\rho_D(\textbf{r})$ is the ionized donor density.  Dirichlet boundary conditions are used to solve Eq.~\ref{eq_poisson}.  $V(\textbf{r})$ is inserted into Eq.~\ref{eq_sch} and the whole procedure is repeated until self-consistency is reached, i.e., the relative variation of the electron density between successive iterations is lower than $10^{-3}$ at any point of the domain.
Finally, the free-carrier linear charge density $n_{\textrm{l}}$ of the NW is calculated as $n_{\textrm{l}}=\int_A n(\textbf{r})d\textbf{r}$, with $A$  the NW cross section.

To assess the impact of the quantum states in these structures, we also compute the linear charge density with a semiclassical approach, where electron states are approximated by plane waves $\psi \propto\exp{\left[i\left(k_x x +k_y y +k_z z\right)\right]}$ with parabolic dispersion, and the wave vectors in the $x$ and $y$ directions are integrated over. The electron density is now obtained from~\cite{datta2005}
\begin{equation}
n_\textsf{sc}(\textbf{r})=2 \left(\frac{m^{\ast} k_B\textrm{T}}{2\pi \hbar^2}\right)^{3/2} \mathcal{F}_{\frac{1}{2}}\left(\frac{E_{\textrm{F}}-E_C(\textbf{r})+eV(\textbf{r}) }{k_B\textrm{T}} \right).
\end{equation}
$n_\textsf{sc}(\textbf{r})$ is then inserted in the Poisson equation~\ref{eq_poisson} in place of $n(\textbf{r})$, and the procedure is iterated like in the fully quantum approach. Clearly, since the solution of (\ref{eq_sch}) is not required, the semiclassical approach is much faster.

\section{System} \label{SP_System} 

We simulated different prototype structures, with the general scheme shown in Fig.~\ref{figure1}. The geometrical parameters of the investigated samples are reported in Tab.~\ref{table1}. Calculations are performed at T = 4~K, and the GaAs and AlGaAs band gaps are taken as 1.43~eV and 1.858~eV, respectively, with the Fermi level $E_{\textrm{F}}$ set at the mid-gap value of GaAs.~\cite{1997handbook} 
Other material parameters are $m^{\ast}(\textrm{GaAs})=0.062$, $\epsilon_r(\textrm{GaAs})=13.18$, $m^{\ast}(\textrm{AlGaAs})=0.092$, and $\epsilon_r(\textrm{AlGaAs})=12.24$.~\cite{1997handbook}

\begin{table}[b]
\centering
\begin{tabular}{ c | c | c | c |c | c | c| c | c}
  \hline
 (nm) &  \ $\Delta_1$ \ & \ $\Delta_2$ \ & \ $\Delta_3$ \ & \ $\Delta_4$ \ & \ $\Delta_5$ \ & \ $\Delta_6$ \ & \ $\Delta_7$ \ & \ $\Delta_8$ \ \\  \hline  
$M_1$ \ & 25  & 25 & 20 & 50 &10 &  0 & 20 & 10 \\  
$S_1$ \ & 25  & 0 &0 & 95 &10 &  0 & 20 & 10 \\  
$S_2$ \ & 25  & 0 &0 & 95 &10 &  0 & 65 & 10 \\  
$M_2$ \ & 25  & 25 & 20 & 50 &10 &  10 & 20 & 10 \\   \hline
\end{tabular}
\caption{ Geometrical  parameters  of the MSNW ($M_1$ and $M_2$) and SSNW ($S_1$ and $S_2$) structures simulated. The widths $\Delta_1$-$\Delta_8$ refer to the labelling in Fig.~\ref{figure1}(a).
\label{table1} }
\end{table}

In the following, we first analyse the multi-shell structure $M_1$, where the effects we want to emphasize are prominent. Then, we show that similar SSNW samples ($S_1$ and $S_2$), or a double-doped MSNW sample, do not present the screening effect. 

\section{MSNW $\mathbf{M_1}$ } \label{SP_M1} 

\begin{figure}[htpb]
  \begin{center} 
    \includegraphics[width=\linewidth]{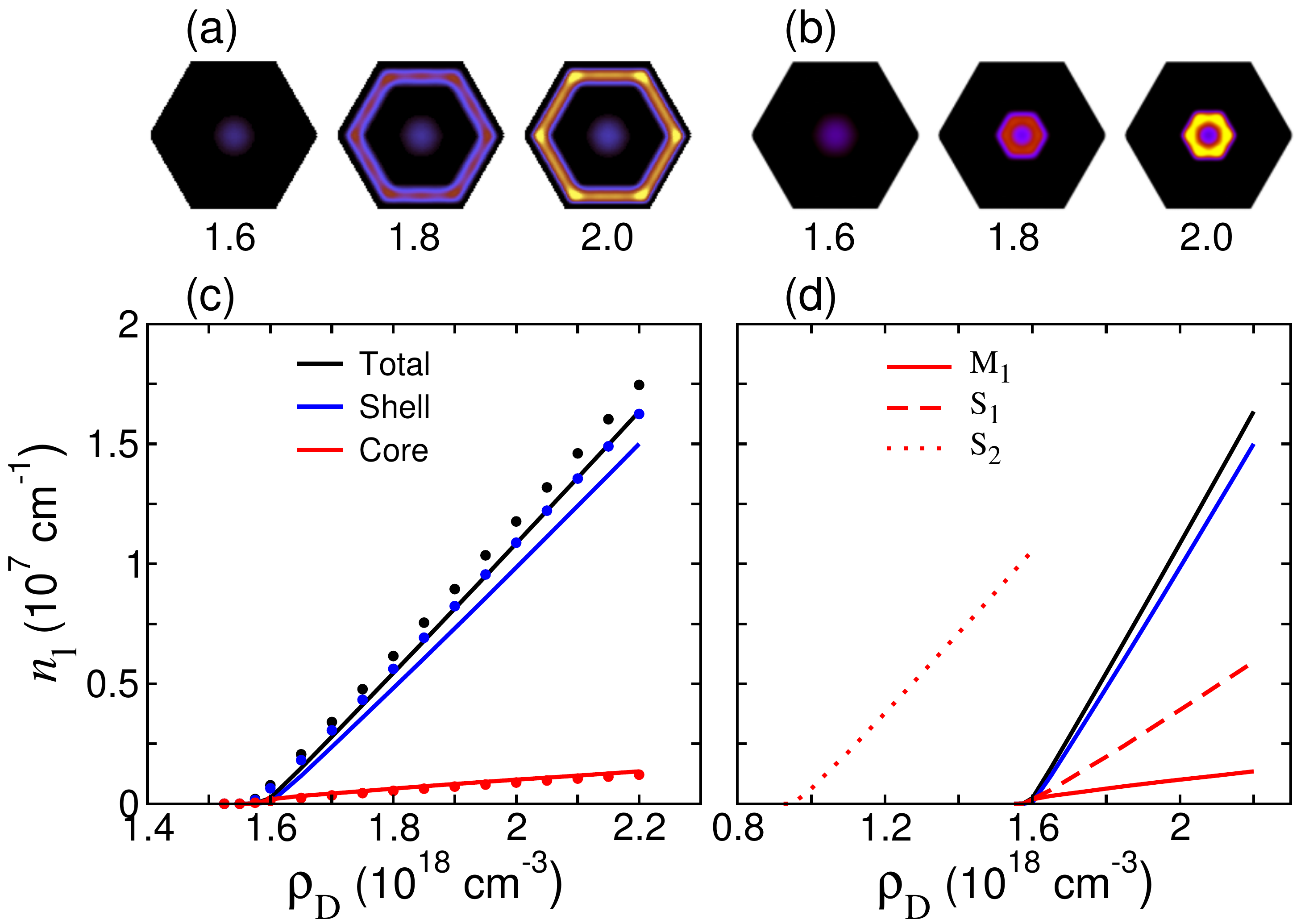}
    \caption{ %Free-charge vs. doping for $M_1$, $S_1$ and $S_2$.
    Electron gas distribution in samples $M_1$ (a) and  $S_1$ (b) at three doping densities. Quantum effects are included.
    (c) Linear charge density $n_{\textrm{l}}$ for sample $M_1$, obtained from the fully quantum approach (solid lines) and the semiclassical approach (dots) \textit{vs} the donor density $\rho_D$ . The three curves report $n_{\textrm{l}}$ in the core (red), in the shell (green) and the total charge density (black). (d) Linear charge density of samples $S_1$ (dashed) and $S_2$ (dotted) compared with $M_1$ (solid lines), where all charge is located in the core. In $M_1$ the charge in the core is much less sensitive to the doping level.
    \label{figure2} }
  \end{center}
\end{figure}

Figure~\ref{figure2}(c) shows the linear free-electron density of structure $M_1$, obtained from the quantum (lines) and semiclassical (dots) approaches \textit{vs} the donor density $\rho_D$, uniformly distributed in a 10-nm-thick layer ($\Delta_6$) in the outer AlGaAs barrier. 
In agreement with other works,~\cite{Tomioka2010,Bertoni2011,Sladek2010}, we find a threshold doping density ($\sim 1.55\times 10^{18}$~cm$^{-3}$) below which no free-electron charge is obtained. Above this threshold, the linear electron densities in the GaAs core and shell increase almost linearly with $\rho_D$.

Charge localization is sensitive to doping concentration.
At low doping $\rho_D=1.6\times 10^{18}$~cm$^{-3}$, just above the threshold (left panel of Fig.~\ref{figure2}(a)), the GaAs core starts to be populated and the charge is distributed with an isotropic circular symmetry.
As $\rho_D$ increases, only a small fraction of the additional free charge accumulates in the core, while most free electrons fill the outer QWT. This is shown, e.g., in the central panel of Fig.~\ref{figure2}(a) at $\rho_D=1.8\times 10^{18}$~cm$^{-3}$, where electrons in the shell form a quasi-uniform sixfold bent 2D gas. At even larger doping, most of the excess free charge still accumulates in the outer shell, but now it tends to form six quasi-1D channels at the QWT bends, due to the larger area available and the sharper bending of the self-consistent field stemming from the repulsive Coulomb potential. This is shown in the right panel of Fig.~\ref{figure2}(a), with $\rho_D=2.0\times 10^{18}$~cm$^{-3}$. 
Being the amount of free charge in the GaAs QWT much larger than the one confined in the core, \emph{it screens the electrostatic field generated by dopants in the central region}. As a consequence, the core electron concentration is weakly affected by the level of doping, as can be gathered from the red curve of Fig.~\ref{figure1}(c).

Figure \ref{figure2} also shows that the results of the fully quantum approach and the semiclassical approch exhibit the same qualitative behaviour, with essentially the same density in the core, and with the density in the GaAs QWT slightly overestimated by the semiclassical approach, due to the lack of quantum confinement effects which shifts the energy subbands upward. 

\section{Comparison with SSNW $\mathbf{S_1}$ and $\mathbf{S_2}$ } \label{SP_S1S2} 

To clarify the effect of the presence of the internal AlGaAs layer $\Delta_2$, we repeated the simulations for two SSNWs with composition and dimension identical to $M_1$, but with $\Delta_2 = 0$.  These two samples, namely $S_1$ and $S_2$, are designed with different position of the doping layer: in $S_1$ ($S_2$) its distance from the outer (inner) GaAs layer is the same as in $M_1$.
The comparison with $M_1$ is shown in Fig.~\ref{figure2}(d). 
As expected, samples $S_1$ and $S_2$ are also populated above a threshold density, which depends on $\Delta_7$, with $n_l$ increasing linearly with doping. Since there is no QWT to accomodate additional charges, charge density in the core can easily reach $\sim 10^7$~cm$^{-1}$ and it is very sensitive to the donor density. Indeed, the conduction band in the central region is significantly bent by the dopants field.
The charge distribution in the cross-section of $S_1$ is illustrated in Fig.~\ref{figure2}(b) for three doping densities. At the higher doping level, the free electrons in the core rearrange into a ring-shaped distribution,\cite{Bertoni2011,SitekPRB2015,BallesterJAP2012} yet keeping a steep linear trend with doping concentration.
To be quantitative, for example, in SSNW $S_2$ a change in $\rho_D$ of $0.2\times 10^{18}~\mbox{cm}^{-3}$
(from $1.0\times 10^{18}$ to $1.2\times 10^{18}~\mbox{cm}^{-3}$)
leads to a variation of $n_l$ of about $500\%$
(from $0.62\times 10^{6}$ to $3.78\times 10^{6}~\mbox{cm}^{-1}$),
while the same doping span
(from $1.8\times 10^{18}$ to $2.0\times 10^{18}~\mbox{cm}^{-3}$) 
in MSNW $M_1$ changes $n_l$ of just $60\%$
(from $0.62\times 10^{6}$ to $1\times 10^{6}~\mbox{cm}^{-1}$).

\section{Effect of shells thickness and doping position} \label{SP_Thickness} 

Having assessed that in MSNWs most of the charge localizes in the QWT and screens the ionized dopants, 
we next investigate the impact of the thickness of the inner barrier, of the well width, and of the doping layer position on the accumulated charge. 
Accordingly, we vary parameters $\Delta_2$, $\Delta_3$, $\Delta_7$ in sample $M_1$, while keeping fixed all other parameters. The total MSNW diameter is varied accordingly.
\begin{figure}[t!]
  \begin{center} 
    \includegraphics[width=\linewidth]{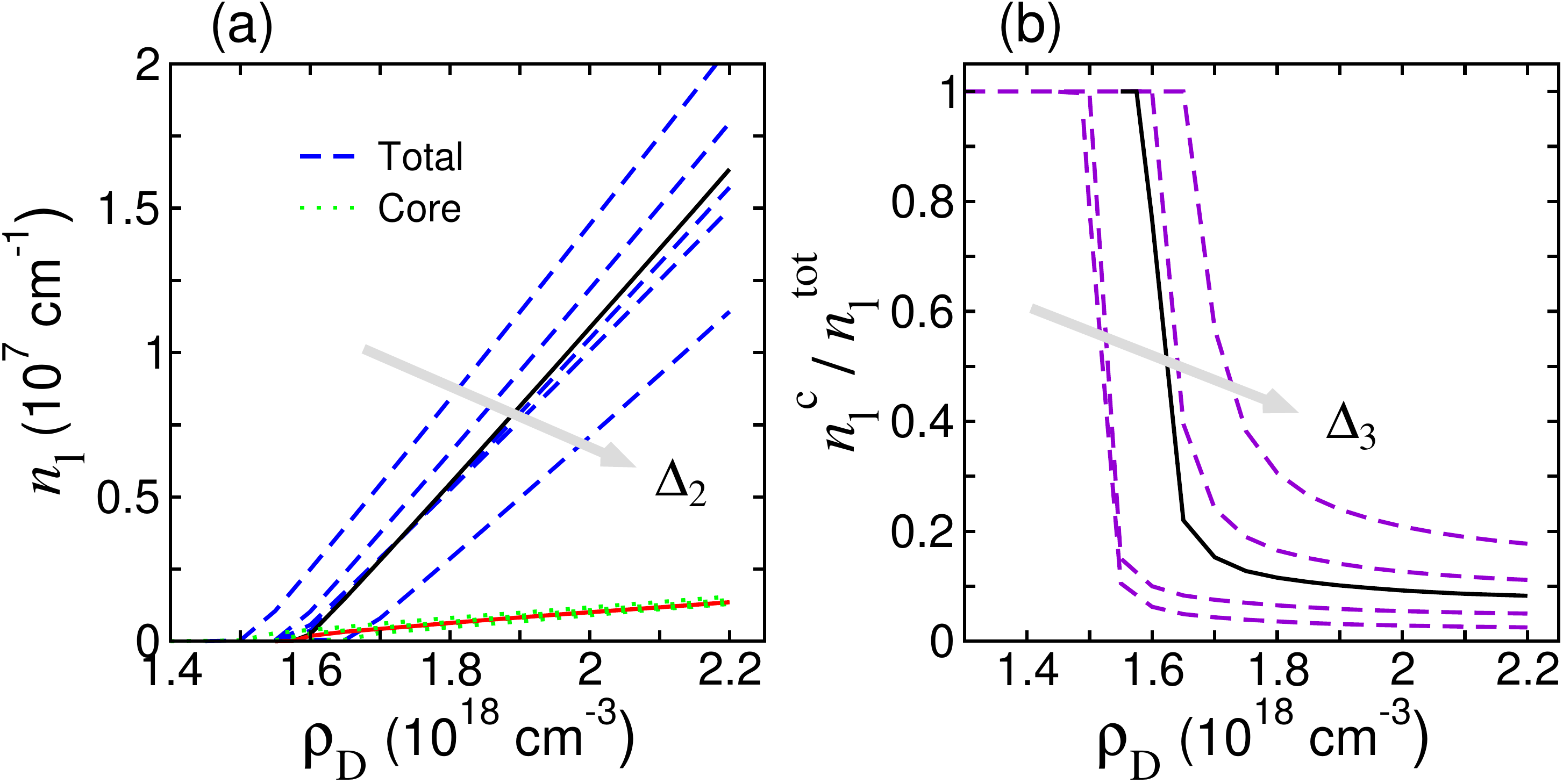}
    \caption{ %Free-charge of $M_1$ with different shells thickness.
    (a) Linear charge density $n_{\textrm{l}}$ in the whole NW (blue dashed lines) and in the core (green dotted lines) \textit{vs} donor density $\rho_D$ for $\Delta_2= 10$, $15$, $20$, $27$, $30$~nm. The gray arrow indicates decreasing values of $\Delta_2$. Vallues for the original MSNW $M_1$, with $\Delta_2=25$~nm is reported with solid lines for easy of comparison. 
    (b) Ratio between the core and the total charge density for $\Delta_3= 10$, $15$, $25$, $30$~nm. The reference structure $M_1$, with $\Delta_3=20$~nm is reported with solid lines for easy of comparison. 
    \label{figure3} }
  \end{center}
\end{figure}

Figure~\ref{figure3}(a) shows that the distance $\Delta_2$ controls the doping threshold by a simple volume effect. 
For given doping density a smaller $\Delta_2$ implies a smaller surface of the hexagonal layer occupied by dopants. Therefore a larger density is needed to accumulate the charge.
Note that the carrier density in the core is practically independent of $\Delta_2$, and the screening effect is always present.

Decreasing the GaAs well $\Delta_3$ has a similar effect on the threshold.
However, the reduction of the screening shell also decreases the amount of charge accumulated in the QWT and the screening effect. Indeed, Fig.~\ref{figure3}(b) shows that the ratio between the core and the total linear density increases for smaller $\Delta_3$.
So, in order for the screening to be effective and, at the same time, to have a significant charge in the core, the GaAs layer thickness must be properly tailored, as in structure $M_1$.

The displacement of the $10$~nm doping layer inserted in $\Delta_4$, has a remarkable effect.
As it moves outwards ($\Delta_7$ decreases) a larger $\rho_D$ is needed to bend the conduction band below the Fermi level. 
This is shown in Fig.~\ref{figure4}(c), with the threshold doping moving to the right as $\Delta_7$ decreases. The effect is stronger for the QWT than for the core, due to their different distance from the doping layer. As a consequence, while the total free-charge density curve becomes less steep, the slope of the core density is unchanged. 
%This means that the screening, as shown in Fig.~\ref{figure4}(c) for $\Delta_7=15$~nm is more effective as the doping get closer to the GaAs shell.

\section{MSNW $\mathbf{M_2}$} 

\begin{figure}
  \begin{center} 
    \includegraphics[width=\linewidth]{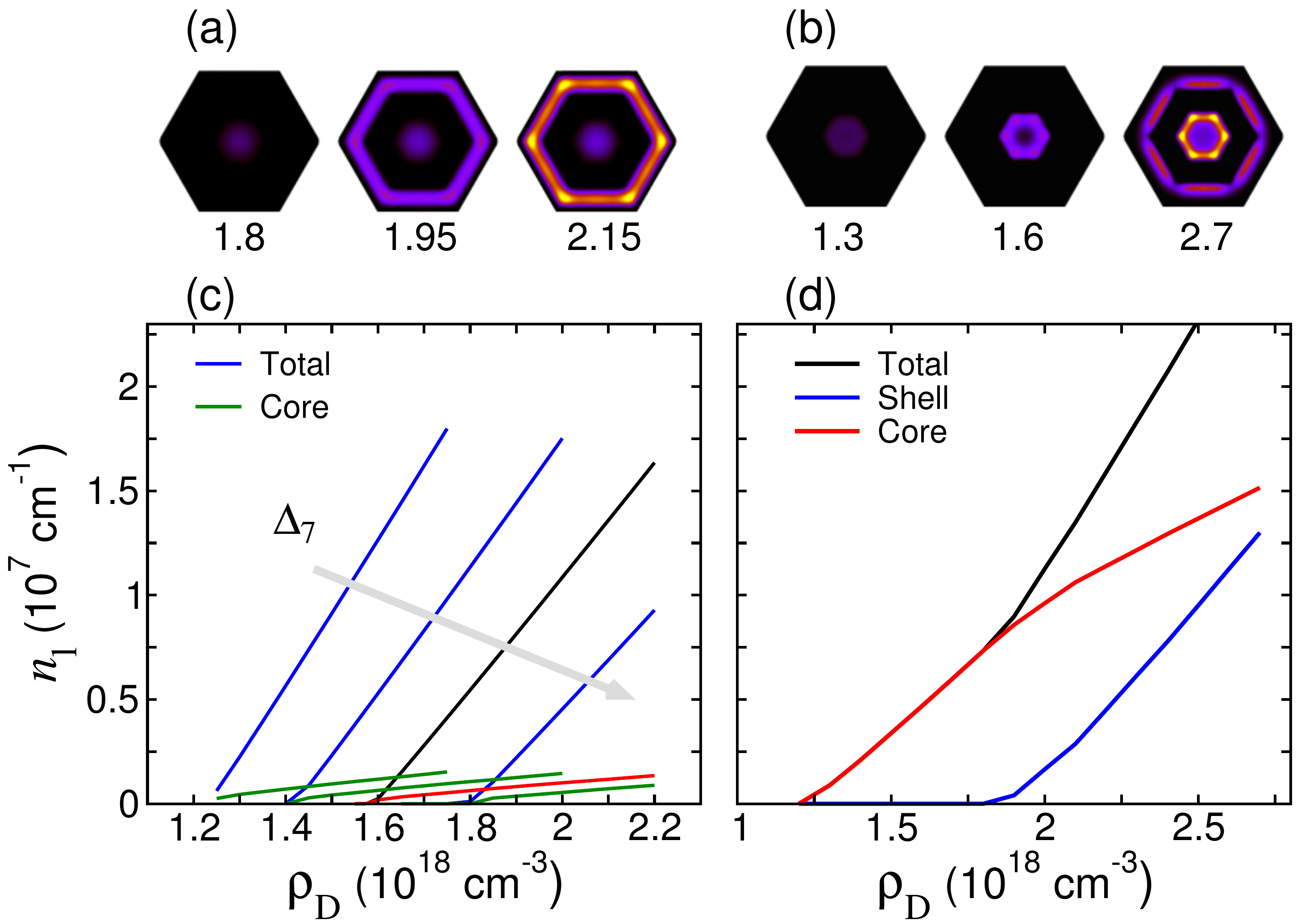}
    \caption{ (a)-(b) Electron gas distribution in $M_1$ with $\Delta_7=15$~nm, and in $M_2$, respectively, at three densities of doping.
    (c) Linear charge density $n_{\textrm{l}}$ against donor density for $M_1$ with $\Delta_7= 15$, $20$, $25$, $30$~nm. The gray arrow indicates decreasing values of $\Delta_7$.
    (d) Linear charge density of $M_2$, with two doping layers.
    \label{figure4}}
  \end{center}
\end{figure}

Finally, we investigate sample $M_2$, where donors are uniformly distributed in \emph{two} $10$~nm layers placed in the two AlGaAs barriers.
The linear free-electron density \textit{vs} $\rho_D$ is shown in Fig.~\ref{figure4}(d) and exhibits two linear regimes, with an abrupt change of slope.
In the lower density regime (from $ 1.22\times 10^{18}$ to $1.8\times 10^{18}$~cm$^{-3}$), free electrons localize only the core region. 
At densities $\rho_D > 1.8\times 10^{18}$ free carriers start to populate the GaAs shell.
The spatial distribution of the electron gas shows peculiar localization patterns, as displayed in Fig.~\ref{figure4}(b).
At low density, the free electron gas is distributed in the center of the core (left panel). As $\rho_D$ increases, a ring-shaped distribution starts to form, with most of  the charge localized on the heterojunction at the edges of the core. 
For larger dopings, when the QWT start to fill, free electrons localize at the facets of the outer GaAs/AlGaAs interface, forming six separated 2DEG strips. 
This also affects the core density, whose slope deviates from the low-doping regime.
A similar emergence of six almost planar electron gases at the hexagon facets has already been obtained in p-doped MSNWs at low density.~\cite{Bertoni2011,Jadc2014,Buscemi_PRB2015}
%This last simulation shows that a doped layer close to the core is detrimental for the screening effect we propose to exploit.

\section{Conclusions}

In conclusion, we showed how the inclusion of a QWT in MSNWs can strongly decrease the effect of doping level uncertainty on the linear free-electron density of the high-mobility core, yet keeping the ability to populate it with a given density of carriers.
We showed numerically that the charge accumulating in the external shell mostly screens the doping electrostatic field in the core for a vast range of shells and doping configurations. 
This property could make MSNWs better candidates for NW-based high-mobility electronic devices.
 
\section*{Acknowledgements}

Numerical simulations were performed at CINECA within the Iscra C project MPL-CSNW. We acknowledge partial financial support from University of Modena and Reggio Emilia, with Grant ``Nano- and emerging materials and systems for sustainable technologies."  MR acknowledges UJI project P1-1B2014-24, MINECO project CTQ2014-60178-P, Beatriu de Pinos program 2014 BP-B 00101.
 
\section*{References}
 
\bibliography{refe_scrCSNW}

\end{document}